# A New Method To Find The Nash Equilibrium Point in Financial Transmission Rights Bidding Problem

Saeed Ahmadian[1], Ramin Farajifijani[2]

[1]S. Ahmadian    Electrical and Computer Engineering department
University of Houston, Houston, Texas 77204
Sahmadian3@uh.edu

[2]R. Farajifijani
Inamori School of Engineering
Alfred University, Alfred, NY 14802
Rf8@alfred.edu

Financial transmission right (FTR) is an important tool and an especially feature for stopping congestion charges in restructured electricity markets. Participants in the transmission market as players are assumed to be a generation company (Gencos) which also take part in an energy market and able to buy their require FTRs. In this regard, there are two types of FTR: obligation or option. There are three main questions which immediately arise for each player who is placed in the market. First, which type of FTR is the best choice? second, how much power is needed to generate by each player and third, how bid prices should be offered. Deciding on these trade-offs is difficult and requires definition of special matrices to measure risk in each possible condition in the transmission market. These matrices include: possibility of flow direction alteration, probable forward and reverse power flow on each line, maximum and minimum offering FTRs and the worst condition of load variation which influence on each player's decision. Based on these matrices, players try to maximize their expected payoffs by taking into account the associated risks. Supposing these matrices are known to respective players, the FTR bidding problem is modeled as a bi-level optimization based on the Nash equilibrium game theory with the upper sub-problem representing player profit maximization and the lower sub-problem representing the optimal solution to the market clearing. An eight-bus system with six players is simulated to verify the proposed method and the obtained results are illustrated the complex interaction between FTR obligation and FTR option bidding strategies. Furthermore, the results are demonstrated to be consistent between the impacts of FTR type, forecasted bid offer of the other players and player's preferred risk levels on FTR bidding strategies. FTR obligation and option, Optimal bidding, Risk concept, New transmission market formulation, Nash equilibrium point, Game theory, transmission market complexities.

## 1   Introduction

With a long-term view of energy markets, there are complementary markets to hedge transmission charges including loss and congestion. Definition of locational marginal price (LMP) based congestion management schemes [1]–[3] is served by competitive transmission markets and organized wholesale electricity markets run by independent operator system [4]–[6]. In other word, a linear's limitation restricts the amount of transmission power flow which leads to diversity in energy nodal prices. Therefore, load buses can pay more than the marginal costs of generation buses. These extra payments motivated the initial idea to form transmission markets. Through this market, participants could hold rights for use of transmission lines through the purchase of financial transmission rights (FTRs). FTR are a sold right which can hedge congestion charges over the constrained transmission routes [7]–[9]. Generally, there are two kinds of FTRs regarding the energy pricing methodology. The first type is called point to point FTR in which the holder must specify a path between two nodes (source and sink) and amount of his required power. This method is executed in many markets such as PJM and New England [10] and [11]. The revenue from FTR holding is calculated as the product of the difference between sources and sinks LMPs and the amount of owned FTR. The latter type is flow gate FTR which is used in decentralized markets. This method is based on zonal marginal pricing in which the FTR owner has the authority to keep special capacity in reserve and planning preference to use links or flow gates between zones [12]. Unlike the first type, owner of flow gate FTR may or may not use it. It is therefore a kind of option. Since this method is not commonly used in of transmission markets [13], it is not considered in this paper. Point to point FTRs (or just FTRs) are generally categorized in two parts; obligation and option [14]–[16]. A FTR obligation holder receives positive profit if the LMP difference between source and sink is positive. In the reverse situation, he would earn negative profit. Therefore, a FTR obligation is a kind liability for its owner. On the other hand, FTR option guarantees positive profit for its owner and when the LMP at the sink is higher than at the source negative profit would not be imposed to the holder. This feature makes FTR option more expensive since risk of negative revenue is neglected. Thus it can be that, given the same price condition for both options an obligation, a FTR option is more preferable. The owner of FTR obligation when the nodal price in the injection point is higher than its value in the withdrawal point must pay an amount. This means that, FTR obligation can be a liability for holders while FTR option will never be in the form of liability [17]. FTR option will be distinct from FTR obligation in several concepts [18]:

1- Simultaneously feasibility test (SFT) is much more difficult for FTR option. After receiving the FTR offers from the market participants, ISO has tendency to maximize revenue from tender on the SFT condition for the congestion and transfer capacity constraints. The aim of SFT is in fact guaranteeing that the gathered congestion charges by ISO from the energy market is not less than the payments to FTR holders under expected and normal congestion performance conditions. The SFT problem by considering FTR obligation is comparable with the security constrained economic dispatch problem. Known algorithms related to the security constrained economic dispatch (SCED) exist; however, because FTR option of the ISOs SFT problem will be required for finding the worst combination of the optimization FTR options for each constraints, it becomes more complex.

2- Only a subset of feasible FTR obligations can exist if they are awarded as FTR options.

Several papers about ISOs FTR option tender market have been discussed. The specific formulas of FTR obligations and options have been presented by using load models dc and ac in [19]. Energy and tender hybrid market has been proposed in [20]. Market participants could accordingly obtain the FTRs along with energy trading. Since energy and transmission markets significantly influence each other, impacts of energy markets on transmission markets and the reverse is regarded in many papers. In [20] and [21] a joint energy and transmission market is analyzed and optimal joint bids and corresponding FTRs are calculated. On the other hand, impact of transmission market on energy participation is discussed in [22] – [25]. In these references, the effect of transmission rights on market power is discussed. It is shown that players in energy market that are located in generating buses can exert market power by purchasing FTRs, while players in demand buses don't have such an opportunity.

In this paper, the focus is on FTR pricing strategies development in the tender markets. During the time of participation in the FTR tender, the bidders will provide the following information for the sale of a specific amount of FTR, which are maximum amount of FTR, bid price and the points of power injection and delivery. The bidders must make their decisions based on the system predicted performance conditions during holding the FTRs. Specifically, they will need that the LMP differences between source points and sinks over a specified FTR route be also considered. Furthermore, powerful competitors and their corresponding pricing information must also be identified. The methods of predicting the electricity price are price simulation method [26], and numerical methods such as artificial neural networks (ANN) [26],[27], time series method [28] and machine learning method [29]. In this paper, it has been assumed that the LMPs have been predicted by using one of the existing methods. Any goal bidding FTR is after maximizing its expected use for holding FTR. Because of the fluctuations of electricity markets, the results of pricing FTR are in fact trade-off between the predicted profit resulting from holding FTR and the risk related to holding a FTR. The risk related to holding a FTR in a bidding object function with risk coefficients has been included. The bidders have information about their objective functions and risk references, but no information on their competitors risk preferences. So, each bidder tries to find the LMPs and their competitor predicted risk references based on historical bidding information and anticipated bidding information under the system performance conditions. It is necessary for the exact modelling of the competitor's special information which can directly affect the incomes of the FTR in the market, to be at the bidder's disposal. Subsequently, the problem of pricing FTR with incomplete information has been formulated as a two surface optimization problem in this paper. The above surface sub-problem states the problem of maximizing the pricing utilities for holding a FTR and the below surface sub-problem also shows the FTR market settlement problem for ISO with the aim of maximizing the incomes obtained from FTR tender while keeping the system convergence.

### 1.1 Problem Statement

In this paper the problem is analyzed from the point of view of a joint participant in energy and transmission markets. It implies that the player as a generator should take part in transmission market as well and guarantees his required power in transmission system. Energy markets are discussed in many different references and are not the subject of this paper but the effect of the energy market have to be regarded as a main point in the decision making process. Modeling an energy market (using DCOPF), parameters which could be estimated are: dispatched power by generators, energy prices and power flow in each line. Using any other procedure to predict these parameters does not influence the major questions facing each participant in transmission market. Briefly they could be categorized as three main parts in the following;

a) Which lines should be chosen as FTR obligation and which ones might be bought as option?

b) How much bid should be presented for each path from each player in transmission market?

c) How much power should be bought?

The first two questions, which constitute definitions of new metrics, are answered in section II and the last question is solved in section III. To sum up these three questions in section IV and objective function with its constraints is presented and using game theory the Nash equilibrium is obtained. Finally, the results are presented in section V on a test case system.

## 2 CHOOSING FTRs AS OBLIGATION OR OPTION AND CORRESPONDING OFFERING PRICES

As declared before, in this paper the problem would be regarded from a Genco's view in the transmission market. This section aims to define the conditions which motivate a player to select a FTR obligation or option. Therefore, for this purpose a decision function is introduced which enables each participant to determine the weather, for each path, an obligation or option should be purchased.

Suppose based on $i^{th}$ player's estimation from energy market, the flow in line between node $l$ and $k$ is $p_{l-k}^{est}$. Therefore Player $i$, has to consider all other player's influence on the line flow between node $l$ and $k$ (this includes himself as well). Mathematically speaking line flow sensitivity for dispatching each generation unit, is a significant parameter for player $i$ and is as below;

$$\psi'^{j}_{l-k} = \frac{\partial p_{l-k}^{est}}{\partial p_j} \qquad (1)$$

In (1) there is a missing issue which has to be considered and that refers to load deviations. In (1) flow sensitivity to generation unit $j$, is stated but the loads supplied by generator $j$ in the event of its dispatch must be considered. Therefore it can be written as;

$$\psi'^{j,d}_{l-k} = \frac{\partial p_{l-k}^{est}}{\partial p_j^d} = A_{l-k,j} - A_{l-k,d} = PTDF_{l-k}^{j-d} \qquad (2)$$

In (2), $\psi'^{j,d}_{l-k}$ shows line $(l-k)$ sensitivity while generator $j$ dispatches to supply load $d$. In fact, it shows impact of generator $j$ on line $l-k$ when it supports load $d$. If there are $N_d$ loads in the network, the simplest way to model the probability of them is to use a probability distribution function. Thus if for load d changing probability is considered $w^d$, it would be clear that;

$$\sum_{d=1}^{N_d} w^d = 1 \qquad (3)$$

Now it can be stated that the expected amount of change in the line between nodes $l$ and $k$ when the load d deviation is $\Delta D_d$ and there is only generator $j$ to supply demand, would be as below; (considering any probability distribution function for load $d$)

$$E(\psi'^{j,d}_{l-k}) = \psi'^{j,d}_{l-k} \Delta D_d w^d \quad (4)$$

But here is an extraordinarily significant issue which is concerned with the energy market and network constraints. When load d changes, based on energy market estimation and by considering limits of transmission lines, output power of each generator can increase or decrease, therefore changes in the network are resulted from these two factors. Mathematically speaking $\Delta D_d = \sum_{j=1}^{N_G} \Delta P_j$ in which $\Delta P_j$ can be positive or negative. If probabilities for increment and reduction in load $d$, are assumed to be $\omega_d^+$, $\omega_d^-$, respectively, and change in generator $j$ output in response of these load deviations, are considered to be $\Delta P_j^{d+}$ and $\Delta P_j^{d-}$ then the expected effect of load d on line between nodes $l$ and $k$ for all condition in network is as follows;

$$\varphi'^d_{l-k} = w^d(l^{d1}_{l-k} + l^{d2}_{l-k}) \quad (5)$$

In (5), $l^{d1}_{l-k}$ and $l^{d2}_{l-k}$ are respectively as below;

$$l^{d1}_{l-k} = \sum_{j=1}^{N_G} \omega_d^+ * \psi'^{j,d}_{l-k} * \Delta P_j^{d+} \quad (6)$$

$$l^{d2}_{l-k} = \sum_{j=1}^{N_G} \omega_d^- * \psi'^{j,d}_{l-k} * \Delta P_j^{d-} \quad (7)$$

It must be noted that $l^{d1}_{l-k}$ and $l^{d2}_{l-k}$ can be positive ($l^{d1+}_{l-k}, l^{d2+}_{l-k}$) or negative ($l^{d1-}_{l-k}, l^{d2-}_{l-k}$). As mentioned in (3) to model load probability the simplest way is to use probability distribution for each load, but this is not the best way to consider the worst situation of transmission market. Indeed if a player in market wants to model the problem with the most conservative view, he must assume that unwanted situations are more probable. In this problem, the situation in which the expected negative flow (reverse flow) for the line between node $l$ and $k$ is increased would be unwanted. Therefore probabilities for each load can be obtained from (8).

$$w^d = \begin{cases} \dfrac{l^{d1-}_{l-k} + l^{d2-}_{l-k}}{\sum_{d=1}^{N_d} l^{d1-}_{l-k} + l^{d2-}_{l-k}} & \text{if } p^{est}_{l-k} > 0 \\[2ex] \dfrac{l^{d1+}_{l-k} + l^{d2+}_{l-k}}{\sum_{d=1}^{N_d} l^{d1+}_{l-k} + l^{d2+}_{l-k}} & \text{if } p^{est}_{l-k} < 0 \end{cases} \quad (8)$$

The expected effect of generator $j$ on line $l-k$ can be positive or negative. Based on these two conditions, two different parameters are introduced as below;

$$FPF_{l-k} = \sum_{d=1}^{N_d} \varphi'^{d+}_{l-k}, \varphi'^{d+}_{l-k} \in \{\varphi'^d_{l-k} \mid \varphi'^d_{l-k} > 0\} \quad (9)$$

$$RPF_{l-k} = \sum_{d=1}^{N_d} \varphi'^{d-}_{l-k}, \varphi'^{d-}_{l-k} \in \{\varphi'^d_{l-k} \mid \varphi'^d_{l-k} < 0\} \quad (10)$$

Coefficients in (9) and (10) are respectively called forward potential flow and reverse potential flow on the line between nodes l and k. The first coefficient has the same direction as the line power flow and the second one is in the opposite direction. These coefficients show the expected amount of power flow in the forward and reverse direction of each given line. Considering these coefficients significantly improves decision making regarding buying FTRs as obligations or options. Suppose the estimated power flow on line $l-k$ is $p_{l-k}^{est}$, then there are two conditions;

a) $p_{l-k}^{est} + FPF_{l-k} > |RPF_{l-k}|$: In this circumstance the main idea coming to mind is that flow direction alteration is not probable. Mathematically, flow increment potential in forward direction is more than flow changing direction potential. Therefore it is more reasonable to buy FTR as obligation, because obligation FTR is less expensive and more preferred to buy.

b) $p_{l-k}^{est} + FPF_{l-k} \leq |RPF_{l-k}|$: In this manner, in spite of previous state, flow return probability is more than its increment in forward direction. It is brilliantly inferred that in this condition risk of purchasing FTR obligation is high and unavoidable. Choosing FTR option may solve risk of earning negative profit.

Based on two abovementioned conditions, chance coefficients are defined as below;

$$\zeta_{l-k}^{f} = \frac{FPF_{l-k}+p_{l-k}^{est}}{FPF_{l-k}+p_{l-k}^{est}+|RPF_{l-k}|} \qquad (11)$$

$$\zeta_{l-k}^{r} = \frac{|RPF_{l-k}|}{FPF_{l-k}+p_{l-k}^{est}+|RPF_{l-k}|} \qquad (12)$$

In (11) $\zeta_{l-k}^{f}$ shows probability that power flow on line $l-k$ keeps its direction, while $\zeta_{l-k}^{r}$ in (12) represents probability of an inversion of the flow direction. For instance suppose that for line $l-k$ these two parameters are; $\zeta_{l-k}^{f} = 0.8, \zeta_{l-k}^{r} = 0.2$. This means that it is 80 percent probable that the power flow remains in its current direction and the probability of inversion is 20 percent. In this condition as a player in transmission market if based on your estimation difference between nodal prices of line $l-k$ is $1 \frac{\$}{MWh}$ then which price do you offer to market for FTR obligation and what about FTR option? To help answer these two questions chance coefficients and estimated nodal price differences are excellent metrics. In other words by using these three factors the questions are answered. Since load centers are integrated far from generation units and Energy flows from generators to load places in the network, it can be assumed that $\zeta_{l-k}^{f} > \zeta_{l-k}^{r}$. Regardless of load or generation places a loss-benefit analysis should be done to Earn maximum limit of prices for each kinds of FTR. Hence two types are FTRs are divided as below:

a) Selecting line $l-k$ as FTR obligation: under this assumption the expected profit by considering $p_{l-k,i}$ MW to buy is as shown in (13);

$$E(profit) = \zeta_{l-k}^{f} p_{l-k,i}(\Delta\lambda_{l-k}^{est} - \pi_{l-k}^{obl}) \qquad (13)$$

And expected loss value is as in (14);

$$E(loss) = \zeta_{l-k}^{r} p_{l-k,i}(\Delta\lambda_{l-k}^{est} - \pi_{l-k}^{obl}) \qquad (14)$$

Player $i$ always foloww the rule that keeps $E(profit) > E(loss)$ and it means that proposed bid must be as below;

$$\pi_{l-k}^{obl} < (2\zeta_{l-k}^{f} - 1)\Delta\lambda_{l-k}^{est} \quad (15)$$

b) Selecting line l-k as FTR option: as before player i expected profit is like (13) but expected loss is as equation (16). Note that in case of choosing FTR option, because it is not associated with negative profit caused by flow direction changing, the only loss is the payment that player i expends to buy FTR option;

$$E(loss) = \zeta_{l-k}^{r} p \ l - k(\pi_{l-k}^{obl}) \quad (16)$$

Therefore offered bid for FTR option is grabbed as (17);

$$\pi_{l-k}^{obl} < \zeta_{l-k}^{f} \Delta\lambda_{l-k}^{est} \quad (17)$$

According to abovementioned points the decision function for player i could be written as below;

$$DF^{i} = [(2\zeta_{l-k}^{f} - 1)\Delta\lambda_{l-k}^{est} - \pi_{l-k}^{obl}]p_{l-k,i}^{obl} + [\zeta_{l-k}^{f}\Delta\lambda_{l-k}^{est} - \pi_{l-k}^{opt}]p_{l-k,i}^{opt} \quad (18)$$

In the next section (18) is modified to consider the required amount of power for player i which is a Genco and is responsible for its power flow in lines.

## 3 ESTIMATION OF REQUIRED POWER

After choosing the FTR type(obligation or option) and their prices, a player (Genco) in transmission market must determine the required power of the selected FTR. To determine this a player in joint transmission and energy markets needs to know about his power flow on each line after clearing the energy market. There are many methods describing how to calculate each generator contribution on lines but in this paper the distribution factors method, as common method, is selected. If $D_{l-k}$ is generation distribution factor for player $i$ then his share on line $l-k$ would be as below [26];

$$p_{l-k,i} = D_{l-k,i} p_{i} \quad (19)$$

To consider other generators' effects on $^{th}$ player's share, a new sensitivity factor is defined as below;

$$\eta_{l-k,i}^{j,d} = \frac{\partial p_{l-k,i}}{\partial p_{j}^{d}} \quad (20)$$

In (20), $\eta_{l-k,i}^{j,d}$ represents sensitivity of $^{th}$ player's share on line $l-k$ with respect to $^{th}$ generator produced power changes to supply load d. Extending equation (20) reaches to (21). It must be regarded that generator $i$ output power is independent from generator $j$ production and consequently the second term in (21) is zero as it can be seen;

$$\eta = \frac{\partial D_{l-k,i}}{\partial p_j^d} p_i + \frac{\partial p_i}{\partial p_j^d} D_{l-k,i}$$
$$\text{Then} \quad \eta = \frac{\partial D_{l-k,i}}{\partial p_j^d} p_i \tag{21}$$

since;

$$D_{l-k,i} = Dl - k, sl + A_{l-k,i} \tag{22}$$

And generalized shift factor $A_{l-k,i}$ is dependent on network configuration and slack bus, not the condition of network, then (21) could be approximated as follow;

$$\eta_{l-k,i}^{j,d} = \frac{D_{l-k,sl}^{it+1} - D_{l-k,sl}^{it}}{\Delta p_j} p_i \tag{23}$$

To calculate , see appendix A.

After calculation of $\eta_{l-k,i}^{j,d}$, expected amount of alteration in share of $i$ th player (or Genco) while generator $j$ changes its output power to supply load $d$ is as below (load deviation is $\Delta p_j$);

$$\chi_{l-k,i}^{j,d} = (\omega_d^+ \eta_{l-k,i}^{j,d} \Delta p_j^{d+} + \omega_d^- \eta_{l-k,i}^{j,d} \Delta p_j^{d-}) \tag{24}$$

In (24) the first term is indexed as $\chi_{l-k,i}^{j,d}$ and the second one is . Because deviation in load d may cause other players in energy market to modify their generation, therefore, expected influence of load d on $i^{th}$ player contribution on line between nodes $l$ and $k$ is represented as follow;

$$v_{l-k,i}^d = w'^d \sum_{j=1}^{N_G} \chi_{l-k,i}^{j,d1} + \chi_{l-k,i}^{d2} \tag{25}$$

The same idea like (8) is posed here to obtain the worst situation for player i. In fact the cases in which the most reverse effect (negative effect) on $i^{th}$ player's contribution are higher, are meant to be more probable. Considering $\chi_{l-k,i}^{d1} = \sum_{j=1}^{N_G} \chi_{l-k,i}^{j,d1}$, $\chi_{l-k,i}^{d2} = \sum_{j=1}^{N_G} \chi_{l-k,i}^{j,d2}$ and $\chi_{l-k,i}^{d1+}$, $\chi_{l-k,i}^{d2+}$ for positive and $\chi_{l-k,i}^{d1-}$, $\chi_{l-k,i}^{d2-}$ for negative values, therefore $w'^d$ in (25) is calculated as below;

$$w'^d = \begin{cases} \frac{\chi_{l-k,i}^{d1-} + \chi_{l-k,i}^{d2-}}{\sum_{d=1}^{N_d} \chi_{l-k,i}^{d1-} + \chi_{l-k,i}^{d2-}} & \text{if} \quad p_{l-k,i} > 0 \\ \frac{\chi_{l-k,i}^{d1+} + \chi_{l-k,i}^{d2+}}{\sum_{d=1}^{N_d} \chi_{l-k,i}^{d1+} + \chi_{l-k,i}^{d2+}} & \text{if} \quad p_{l-k,i} < 0 \end{cases} \tag{26}$$

Again potential coefficients are applied here but this time in form of forward contribution potential (FCP) and reverse contribution potential (RCP) coefficients;

$$FCP^i_{l-k} = \sum_{d=1}^{n_d} v^{d+}_{l-k,i} \quad , v^{d+}_{l-k,i} \in \{v^j_{l-k,i}|v^j_{l-k,i} > 0\} \tag{27}$$

$$RCP^i_{l-k} = \sum_{d=1}^{n_d} v^{d-}_{l-k,i} \quad , v^{d-}_{l-k,i} \in \{v^d_{l-k,i}|v^d_{l-k,i} > 0\} \tag{28}$$

In (27), $FCP^i_{l-k}$ represents expected flow (MW) that intensifies share of player $i$, while $RCP^i_{l-k}$ in (28) is expected amount of power (mw) which reduces player $i$ contribution on line $l-k$. Using these two coefficients makes two possible states. These possibilities are respectively as follows;

1) $FCP^i_{l-k} > |RCP^i_{l-k}|$: In this case, it can be inferred that chance of increment in share of player $i$ is more than its decrement. Therefore if he tends to buy $p_{l-k,i}$ as his required power, he is exposed to risk that his share would be more than $p_{l-k,i}$ and consequently he has to pay extra cost equal to (29) to compensate its probable extra power flow on line $l-k$.

$$(FCP^i_{l-k} - |RCP^i_{l-k}|)\Delta\lambda^{est}_{l-k} \tag{29}$$

thus in this condition its total profit would be as (30);

$$R_i = \left(p_{l-k,i} - (FCP^i_{l-k} - |RCP^i_{l-k}|)\right) \times \Delta\lambda^{est}_{l-k} - p_{l-k,i}\pi_{l-k} \tag{30}$$

In (30), $\pi_{l-k}$ stand for either FTR obligation or FTR option price.

2) $FCP^i_{l-k} > |RCP^i_{l-k}|$: This situation approves that the amount of $i^{th}$ player's share, possibly decreases and this means that he has bought power more than he needs. Under this condition he is entailed the rule of use it or loss it since there is no secondary market to sell his extra amount of power. In other word, he pays extra money for the power that he doesn't use. The imposed cost to him is stated in (31).

$$(|RCP^i_{l-k}| - FCP^i_{l-k})\pi_{l-k} \tag{31}$$

Based on (31), in this circumstance his profit is;

$$R_i = p_{l-k,i}\Delta\lambda^{est}_{l-k} - [p_{l-k,i} + |RCP^i_{l-k}| - FCP^i_{l-k}]\pi_{l-k} \tag{32}$$

According to abovementioned points, $i^{th}$ player's FTR on line $l-k$ would be definitely within below interval;

$$p^{est}_{l-k,i} - |RCP^i_{l-k}| < FTR_{l-k,i} < p^{est}_{l-k,i} + FCP^i_{l-k} \tag{33}$$

In (33), $p^{est}_{l-k,i}$ represents the estimated amount of player's share on line $l-k$, gained by energy market simulation. At last regarding both possible states, by combining (30) and (32) the final profit function is presented in (34).

$$R_i^{final} = \left(FTR_{l-k,i} - max\left(0, FCP_{l-k}^i - |RCP_{l-k}^i|\right)\right)$$
$$\Delta\lambda_{l-k}^{est} - \left(FTR_{l-k,i} - min\left(0, FCP_{l-k}^i - |RCP_{l-k}^i|\right)\right)\pi_{l-k} \quad (34)$$

## 4  NEW FORMULATION FOR TRANSMISSION MARKET

So far the three main questions posed in part I section A, have been answered separately. In this section the FTR bidding problem as an optimization problem with a new formulation is introduced. It should be noted that the optimization variables are respectively; a set of binary variables representing the choice of FTR obligation or option for each path, the amount of power required for each line and finally the proposed price for the required FTR. Before presenting the objective function, two parameters are defined as positive FTR in (35) and negative FTR in (36).

$$FTR_{l-k,i}^+ = FTR_{l-k,i} - max\left(0, FCP_{l-k}^i - |RCP_{l-k}^i|\right) \quad (35)$$

$$FTR_{l-k,i}^- = FTR_{l-k,i} - min\left(0, FCP_{l-k}^i - |RCP_{l-k}^i|\right) \quad (36)$$

Based on (18) and (34) objective function for this problem its associated constraints are presented in (37) to (40).

$$obj = \\ \left[max\left(0, \frac{\xi_{l-k}^f - \xi_{l-k}^r}{|\xi_{l-k}^f - \xi_{l-k}^r|}\right) \times \right. \\ \left.\left((2\xi_{l-k}^f - 1)\Delta\lambda_{l-k}^{est}FTR_{l-k,i}^{+obl}\right) - \pi_{l-k}^{obl}FTR_{l-k,i}^{-obl}\right] \\ + \left[\xi_{l-k}^f \Delta\lambda_{l-k}^{est}FTR_{l-k,i}^{+opt} - \pi_{l-k}^{opt}FTR_{l-k,i}^{-opt}\right] \quad (37)$$

Subject to;

$$p_{l-k,i}^{est} - |RCP_{l-k}^i| < FTR_{l-k,i} < p_{l-k,i}^{est} + FCP_{l-k}^i \quad (38)$$

$$\rho_{l-k}^{obl,base} < \rho_{l-k}^{obl} < (2\zeta_{l-k}^f - 1)\Delta\lambda_{l-k}^{est} \quad (39)$$

$$(2\zeta_{l-k}^f - 1)\Delta\lambda_{l-k}^{est} < \rho_{l-k}^{opt} < \zeta_{l-k}^f \Delta\lambda_{l-k}^{est} \quad (40)$$

### 4.1  ISO Problem

To solve the optimization problem in (37)-(40), first it is needed to solve ISO market clearing price to obtain allocated FTRs and corresponding prices which is mentioned in [30] as below;

$$max \sum_{i=1}^{N_p^{obl}} \rho_i^{obl}FTR_i^{obl} + \sum_{i=1}^{N_p^{opt}} \rho_i^{opt}FTR_i^{opt} \quad (41)$$

S.T.

$$p_{l-k} = \sum_{i=1}^{N_p^{obl}} M_i FTR_i^{obl} + \sum_{i=1}^{N_p^{opt}} max(0, M_i) FTR_i^{opt} \qquad (42)$$

$$FTR_i^{min,obl} < FTR_i^{obl} < FTR_i^{max,obl} \quad i = 1, \ldots, N_p^{obl} \qquad (43)$$

$$FTR_i^{min,opt} < FTR_i^{opt} < FTR_i^{max,opt} \quad i = 1, \ldots, N_p^{opt} \qquad (44)$$

Therefore, each player has to solve ISO problem at first and it means that players have to deal with a bi-level optimization problem as below;

$$\begin{aligned} & max \quad obj_i \\ & S.T. \\ & \rho_{l-k}^{obl,min} < \rho_{l-k}^{obl} < \rho_{l-k}^{obl,max} \\ & \rho_{l-k}^{opt,min} < \rho_{l-k}^{opt} < \rho_{l-k}^{opt,max} \\ & ISO Problem \end{aligned} \qquad (45)$$

### 4.2 Game Theory

To solve the bi-level problem in (45), there are methods such as branch and bound, steepest decent direction and heuristic techniques which can be applied. In this paper by using Karush-Kuhn-Tucker (KKT) optimality conditions the problem is changed to a single level optimization (see appendix B) and all player's bids are obtained by solving that. In fact all players are assumed to compete in a game theory problem and the by solving the below equations all bids are computed simultaneously. However one can solve the problem serially. It means that player i by considering initial bids for other players, can update his bid until there are no more changes in his bid. Sequentially after that the player j would follow this procedure and this would be repeated by all players until anyone doesn't change his bid. But as it mentioned, here all bids, are obtained in parallel by executing the following optimization problem;

$$\begin{aligned} max \quad & \sum_{i=1}^{N_p} \sum_{j=1}^{N_{path}} \\ & max\left(0, \frac{\zeta_j^f - \zeta_j^r}{|\zeta_j^f - \zeta_j^r|}\right) ((2\zeta_j^f - 1) \Delta \lambda_j^{est} FTR_{j,i}^{+obl} - \\ & \pi_j^{obl} FTR_{j,i}^{-obl}) + (\zeta_j^f \Delta \lambda_j^{est} FTR + j, i^{+opt} - \\ & \pi_j^{opt} FTR_{j,i}^{-opt}) \end{aligned} \qquad (46)$$

Subject to quality constraint of ISO problem which is:

$$\begin{aligned} p_j = \sum_{i=1}^{N_p^{obl}} M_i^j FTR_{j,i}^{obl} + \sum_{i=1}^{N_p^{opt}} max(0, M_i) FTR_{j,i}^{opt} \\ j = 1, \ldots, N_{path} \end{aligned} \qquad (47)$$

KKT Optimality consition for $FTR_i^{obl} < FTR_i^{max,obl}$;

$$FTR_{j,i}^{obl} + s_{j,i}^{obl} = p_{j,i}^{est} + FCP_j^i \qquad (48)$$

$$s_{j,i}^{obl} \mu_{j,i}^{+obl} = 0 \qquad (49)$$

KKT Optimality condition for $FTR_i^{opt} < FTR_i^{max,opt}$;

$$FTR_{j,i}^{opt} + s_{j,i}^{opt} = p_{j,i}^{est} + FCP_j^i \qquad (50)$$

$$s_{j,i}^{opt} \mu_{j,i}^{+opt} = 0 \qquad (51)$$

Where $s_{j,i}^{obl}$ and $s_{j,i}^{opt}$ are slack variables. KKT optimality condition for $FTR_i^{min,obl} < FTR_i^{obl}$ and $FTR_i^{min,opt} < FTR_i^{opt}$ respectively are;

$$\mu_{j,i}^{-obl}(FTR_{j,i}^{obl} + |RCP_j^i| - p_{j,i}^{est}) = 0 \qquad (52)$$

$$\mu_{j,i}^{-opt}(FTR_{j,i}^{opt} + |RCP_j^i| - p_{j,i}^{est}) = 0 \qquad (53)$$

Lagrangian function for ISO problem is presented in (54) as below;

$$\begin{aligned} L = &\sum_{i=1}^{N_p^{obl}} \rho_i^{obl} FTR_i^{obl} + \sum_{i=1}^{N_p^{opt}} \rho_i^{opt} FTR_i^{opt} + \\ &\mu_j^{eq}(p_{l-k} - (\sum_{i=1}^{N_p^{obl}} M_i FTR_i^{obl} + \sum_{i=1}^{N_p^{opt}} max(0, M_i) FTR_i^{opt})) \\ &+\mu_{j,i}^{+obl}(FTR_{j,i}^{obl} - p_{j,i}^{est} + FCP_j^i) + \\ &\mu_{j,i}^{+opt}(FTR_{j,i}^{opt} - p_{j,i}^{est} + FCP_j^i) + \\ &\mu_{j,i}^{-obl}(p_{j,i}^{est} - |RCP_j^i| - FTR_{j,i}^{-opt}) + \\ &\mu_{j,i}^{-opt}(p_{j,i}^{est} - |RCP_j^i| - FTR_{j,i}^{opt}) \end{aligned} \qquad (54)$$

Optimality condition for $\frac{\partial L}{\partial FTR_{j,i}^{obl}}$ and $\frac{\partial L}{\partial FTR_{j,i}^{opt}}$ implies that;

$$\mu_{j,i}^{+obl} + \rho_{j,i}^{obl} - M_i^j \mu_j^{eq} - \mu_{j,i}^{-obl} = 0 \qquad (55)$$

$$\mu_{j,i}^{+opt} + \rho_{j,i}^{opt} - max(0, M_i) \mu_j^{eq} - \mu_{j,i}^{-opt} = 0 \qquad (56)$$

Finally bid constraints respectively are;

$$\rho_j^{obl,bas} \leq \rho_{j,i}^{obl} \leq (2\zeta_j^f - 1)\Delta\lambda_j^{est} \qquad (57)$$

$$(2\zeta_j^f - 1)\Delta\lambda_j^{est} \leq \rho_{j,i}^{opt} \leq \zeta_j^f \Delta\lambda_j^{est} \qquad (58)$$

To solve above mentioned optimization every heuristic and Meta-heuristic method can be applied. In this paper interior-point algorithm is implemented [31]. The answer of optimization problem is Nash equilibrium in which optimal bids, amount of FTRs and their types (option or obligation) are obtained. In next section results for proposed method on test network is presented.

## 5 Results

To verify the proposed method an 8 bus network is considered here. Network data can be found in [32]. It should be noted that the capacity of line 10 is restricted to 9 MW. Fig. 1, shows the network with six players taking part in the transmission market (remember that the players are assumed to participate in a joint energy and transmission market).

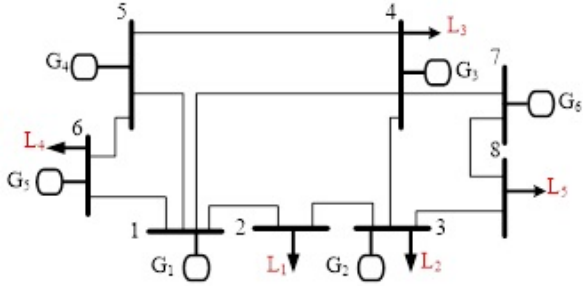

Figure 1: Test system with 8 buses and 6 generators

To model the competition between six players, five paths are considered here, including line 1, 2, 8, 10 and 11. Load variation is considered to be 10 percent for each one. Values for $\zeta_j^f$ are listed in table 1. As it can be seen, power flow on line 8 has the most potential to change among rest of paths and flow on line 10 is the least probable to alter. In other word line 8 is not suitable for buying FTR obligation while line 10 seems to be a safe option to buy FTR obligation.

Table 1: Values of $\zeta_j^f$ for Selected Paths

| Line | 1 | 2 | 8 | 10 | 11 |
|---|---|---|---|---|---|
| $\zeta_j^f$ | 0.9113 | 0.7161 | 0.4947 | 0.9139 | 0.802 |

Table 2 shows values of FCP and RCP (MW) for each player on selected paths. The table shows that in line 1, the greatest change is related to player 1 in the reverse direction. In fact, table 2 gives an initial view to each player about his required share. For example according to (18), first player's contribution on line 1 is 9.7966 MW but by considering table 2, he could possibly reduce his share to 7.9503 MW. Clearly, if he buys all his contribution (9.7966 MW) in transmission market, he would be required to $'use \quad it \quad or \quad lose \quad it'$ because his share would be less than 9.7966 actually based on table 2. This is reverse for him on line 8. Because based on table 2, his share would increase about 3.1742 MW and this declares that he has to buy more than his actual share.

Table 2: FCP and RCP of Players for Each Line

| Line | 1 | 2 | 8 | 10 | 11 |
|---|---|---|---|---|---|
| $FCP_j^1$ | 0 | 0 | 3.1742 | 0 | 1.7990 |
| $RCP_j^1$ | -7.9503 | -6.1861 | 0 | -3.1254 | 0 |
| $FCP_j^2$ | 0.2801 | 0.0649 | 0.2137 | 0.1177 | 0.5559 |
| $RCP_j^2$ | 0 | 0 | 0 | 0 | 0 |
| $FCP_j^3$ | 0.2210 | 0.3604 | 0.1889 | 0.3874 | 0.0093 |
| $RCP_j^3$ | -0.2526 | -0.3160 | -0.1656 | -0.4419 | -0.0081 |
| $FCP_j^4$ | 2.8120 | 1.8102 | 0 | 1.3634 | 0 |
| $RCP_j^4$ | 0 | 0 | -0.1941 | 0 | -0.6388 |
| $FCP_j^5$ | 3.8437 | 2.7283 | 4.6066 | 1.7084 | 0 |
| $RCP_j^5$ | 0 | 0 | 0 | 0 | -0.8964 |
| $FCP_j^6$ | 0.0365 | 0.2022 | 0.1424 | 0.3033 | 0.1462 |
| $RCP_j^6$ | -0.0755 | -0.0976 | -0.0688 | -0.6282 | -0.3027 |

To simulate players' behaviors in the transmission market as a game and choosing their decisions about FTR obligation or option, two states are considered. These two states are the worst conditions that each player may encounter. First it is supposed that all players select all paths as obligation. in this condition, each participant who bids more would be in the set of winners. Second, it is assumed that all players offer bids as option. When all players take a special path as option, it means that, market clearing price would be more than same situation selecting as obligation. Better to say if from all 6 generators on the transmission market, only one generator bids as obligation, his chance to earn FTR is least, since his proposed bid is clearly less than others. Table 3 shows players' profits ($/MWh) in these two situations. Nash equilibrium points for each player on each path, are bolted. Regarding output results, it is better for player 1 to buy FTR on line 1 as option because his profit when he chooses option is more than obligation. This is true for line 2 and 8 also. The same analysis can be done for other players.

Optimal bids for players considering each of two above mentioned states are listed in table 4. Proposed bids for options are more than bids for obligations and this is because of FTR option inherence which is more expensive and less risky based on (44). Market clearing prices for both option and obligation FTRs are also collected in table 5.

Allocated FTRs in each state are drawn in table 6. Since flow direction of line 8 possibly changes, no one would buy FTR on line 8 as obligation. Contribution of Player 1 in line 1 and 2 when he selects them as FTR option has increased. Looking at table 4, declares that despite of bid increment for player 1 on line 1 and 2 (consequently MCP increment for option), increment in his FTR option and reduction in risk makes him to pick these two lines as option (based on table 3). But on line 11, allocated FTR to him by choosing obligation is more and since market clearing price in this case is lower (table 5) by choosing obligation, his profit would be more.

Table 3: Players' Profit When Choosing FTRs as Obligation or Option $ (*p:profit)

| Line | 1 | 2 | 8 | 10 | 11 |
|---|---|---|---|---|---|
| $*p^{obl,1}$ | 592.0559 | 53.8029 | 0 | **364.1888** | **5.1286** |
| $p^{opt,1}$ | **694.46** | **64.077** | **0.6426** | 315.21 | 1.0882 |

|  | 00 | 1 |  | 94 |  |
|---|---|---|---|---|---|
| $p^{obl,2}$ | **163.8827** | **4.5442** | 0 | 86.6924 | 79.7370 |
| $p^{opt,2}$ | 158.7509 | 4.5193 | **1.4291** | **86.6924** | **81.6740** |
| $p^{obl,3}$ | **92.6898** | **15.4569** | 0 | 204.5260 | **0.8149** |
| $p^{opt,3}$ | 90.0571 | 15.3798 | **0.9488** | 204.5260 | 0.08142 |
| $p^{obl,4}$ | **-53.5526** | **-16.5900** | 0 | -131.447 | 0 |
| $p^{opt,4}$ | -215.340 | -16.5989 | **0.0115** | -82.4536 | **4.4067** |
| $p^{obl,5}$ | -294.344 | **-14.1617** | 0 | -60.0880 | **10.1666** |
| $p^{opt,5}$ | **-227.694** | -25.0169 | -3.7353 | -76.5785 | 7.7633 |
| $p^{obl,6}$ | **27.9203** | **7.9899** | 0 | 292.3031 | **27.4296** |
| $p^{opt,6}$ | 27.0644 | 7.9496 | **0.5910** | 308.7688 | 27.4039 |

Optimal bids for players considering each of two above mentioned states are listed in table 4. Proposed bids for options are more than bids for obligations and this is because of FTR option inherence which is more expensive and less risky based on (44). Market clearing prices for both option and obligation FTRs are also collected in table 5.

Table 4: Optimal Bids for Each Player When Choosing FTRs as Option or Obligation

| Line | 1 | 2 | 8 | 10 | 11 |
|---|---|---|---|---|---|
| $Bid^{obl,1}$ | 15.6259 | 3.8579 | 0 | 19.4676 | 9.1501 |
| $Bid^{opt,1}$ | 73.0018 | 7.6313 | 0.5602 | 91.9859 | 16.6840 |
| $Bid^{obl,2}$ | 11.0920 | 3.4378 | 0 | 14.5577 | 9.1501 |
| $Bid^{opt,2}$ | 71.9924 | 7.4518 | 0.5602 | 91.9623 | 16.6840 |
| $Bid^{obl,3}$ | 15.6259 | 3.3938 | 0 | 19.4679 | 9.1503 |
| $Bid^{opt,3}$ | 73.0018 | 7.6486 | 0.5602 | 91.9859 | 16.6840 |
| $Bid^{obl,4}$ | 15.6259 | 3.8579 | 0 | 19.4676 | 9.1501 |
| $Bid^{opt,4}$ | 73.0018 | 7.6313 | 0.5602 | 91.9859 | 16.6840 |
| $Bid^{obl,5}$ | 17.6332 | 3.8579 | 0 | 19.4676 | 9.1501 |
| $Bid^{opt,5}$ | 73.0018 | 7.6313 | 0.5602 | 91.9859 | 16.6840 |
| $Bid^{obl,6}$ | 15.6260 | 3.4351 | 0 | 19.4676 | 11.2998 |
| $Bid^{opt,6}$ | 73.0018 | 7.5326 | 0.5602 | 91.9859 | 16.6840 |
| | | | | | |

Finally, Fig. 2, shows final profit and players' decisions in selecting FTRs obligation or option based on table 3. It implies that players generally tends to bid FTR obligation on lines 1, 2 and 11 while it is reverse for lines 8 and 10.

Table 5: MCPs for FTR Obligations and Options $/MW

| Line | 1 | 2 | 8 | 10 | 11 |
|---|---|---|---|---|---|
| $MCP^{obl,1}$ | 15.6259 | 3.8579 | 0 | 19.4676 | 9.1501 |
| $MCP^{opt,1}$ | 73.0018 | 7.6313 | 0.5602 | 91.9859 | 16.6840 |
| | | | | | |

Table 6: FTRs for Obligation and Option Each State

| Line | 1 | 2 | 8 | 10 | 11 |
|---|---|---|---|---|---|
| $FTR^{obl,1}$ | 7.7315 | 5.8676 | 0 | 3.7775 | 2.0722 |
| $FTR^{opt,1}$ | 8.9491 | 7.0524 | 3.9667 | 3.2696 | 1.8584 |
| $FTR^{obl,2}$ | 2.4201 | 0.5604 | 0 | 1.0169 | 4.8036 |
| $FTR^{opt,2}$ | 2.4201 | 0.5604 | 1.9761 | 1.0169 | 4.9090 |
| $FTR^{obl,3}$ | 1.2104 | 1.7301 | 0 | 2.1214 | 0.0446 |
| $FTR^{opt,3}$ | 1.2104 | 1.7301 | 1.1933 | 2.1214 | 0.0446 |
| $FTR^{obl,4}$ | 2.1127 | 0.0010 | 0 | 0 | 0 |
| $FTR^{opt,4}$ | 0 | 0 | 0.0142 | 0.5082 | 0.2354 |
| $FTR^{obl,5}$ | 0 | 1.1838 | 0 | 1.0851 | 0.5416 |
| $FTR^{opt,5}$ | 0.8952 | 0 | 0 | 0.9141 | 0.4146 |
| $FTR^{obl,6}$ | 0.3646 | 0.9759 | 0 | 3.0319 | 1.4611 |
| $FTR^{opt,6}$ | 0.3646 | 0.9759 | 0.8025 | 3.2027 | 1.4611 |

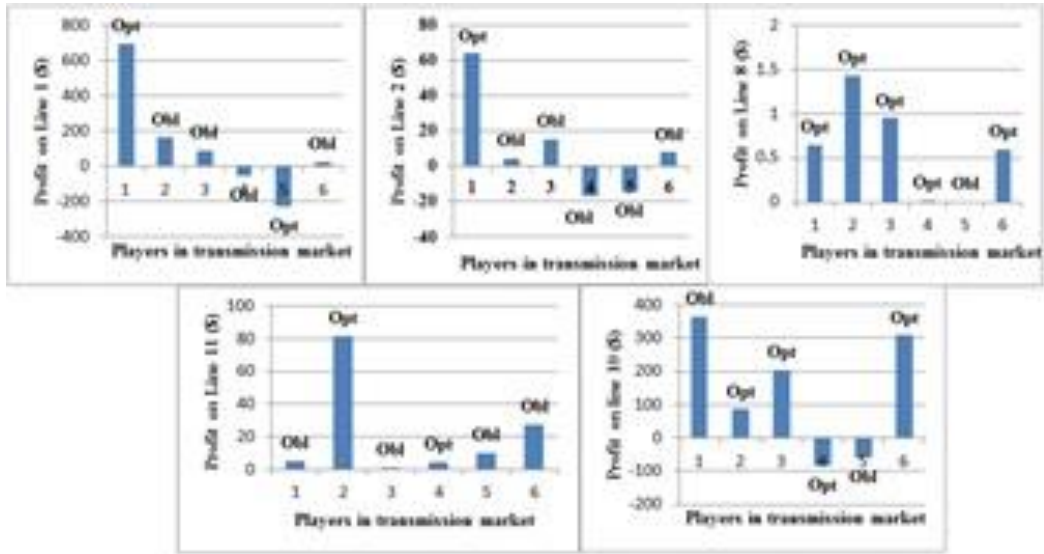

Figure 2: Players' Profits and Their Decisions To Buy FTR Option or Obligatoin on Each Line

## 6 Conclusion

The transmission market is designed to help its participants to hedge congestion charges. The most common instruments applied in this market are FTRs. FTR is an effective tool for hedging congestion charges in the electricity markets. Due to ambiguities and the corresponding risks of participation in market and the decision making process, determining the optimum purchase of FTR is difficult for players. A bi-level model has been proposed for calculating the optimum pricing strategy in the FTR auction markets. The upper-level sub-problem has been modelled separately as a problem for maximizing the profit based on risk for individual bidders. The lower-level sub-problem is also implemented based on a FTR market settlement problem for ISO. The bi-level problem has been solved based on sensitivity functions. By means of new metrics and parameters such as maximum and minimum amount of FTRs, offering prices and chance coefficients, the three questions that each player faces in the market are answered by the evolutionary game theory approach. Eventually, a test network is used to implement the proposed algorithm and the Nash equilibrium is obtained. An eight bus system has been simulated for investigating the performance of the proposed method. The obtained results showed the pricing differences between FTR obligation and option. The paper has shown that the congestion constraints might cause the pricing tender to be more attractive to certain bidders. Furthermore, the obtained results showed that an exact prediction of LMP difference and the adequate risk efficiency could play a significant role in FTR pricing and the bidders' profits.

# 7 Appendix

## 7.1 Appendix A

Suppose that $D^{it}_{l-k,sl}$ is the initial generalized generation factor of slack bus at balance equilibrium point. Based on [21] it can be written as below;

$$D^{it}_{l-k,sl} = \frac{P^{it}_{l-k} - \sum_{i=1, i \neq sl}^{N_G} A_{l-k,i} P_i}{\sum_{i=1}^{N_G} P_i} \tag{59}$$

When generator $j$ dispatches to supply load $d$, power flow on line between nodes $l$ and $k$ definitely faces one of the below conditions;

1) If $[(A_{l-k,j} - A_{l-k,d})\Delta P_j] - [P^{max}_{l-k} - P^{it}_{l-k}] < 0$ then;

$$P^{it+1}_{l-k} = P^{it}_{l-k} + [(A_{l-k,j} - A_{l-k,d})\Delta P_j] \tag{60}$$

2) If $[(A_{l-k,j} - A_{l-k,d})\Delta P_j] - [P^{max}_{l-k} - P^{it}_{l-k}] > 0$ then;

$$P^{it+1}_{l-k} = P^{max}_{l-k} \tag{61}$$

Base on each of the above conditions $D^{it+1}_{l-k,sl}$ can be calculated. Therefore, for the both of them, the conditions are analyzed.

Suppose that first condition is satisfied, then it can be written;

$$D^{it+1}_{l-k,sl} = \frac{1}{\sum_{i=1}^{N_G} P_i + \Delta P_j} (P^{it}_{l-k} + ((A_{l-k,j} - A_{l-k,d})\Delta P_j) - \sum_{i=1, i \neq sl}^{N_G} A_{l-k,i} P_i - A_{l-k,j}\Delta P_j) \tag{62}$$

Substituting (59) into (62) results in;

$$D^{it+1}_{l-k,sl} = \frac{D^{it}_{l-k,sl} \sum_{i=1}^{N_G} P_i - A_{l-k,d}\Delta P_j}{\sum_{i=1}^{N_G} P_i + \Delta P_j} \tag{63}$$

Therefore the difference between $D^{it+1}_{l-k,sl}$ and $D^{it}_{l-k,sl}$ easily would be obtained.

2) Consider that second condition is true then we have;

$$D^{it+1}_{l-k,sl} = \frac{P^{max}_{l-k} \sum_{i=1, i \neq sl}^{N_G} A_{l-k,i} P_i - A_{l-k,j}\Delta P_j}{\sum_{i=1}^{N_G} P_i + \Delta P_j} \tag{64}$$

After some manipulations;

$$D^{it+1}_{l-k,sl} = \frac{D^{it}_{l-k,sl} \sum_{i=1}^{N_G} P_i + (P^{max}_{l-k} - P^{it}_{l-k} - A_{l-k,j}\Delta P_j)}{\sum_{i=1}^{N_G} P_i + \Delta P_j} \tag{65}$$

In this situation like the first condition the difference is earned apparently.

## 7.2 Appendix B

To solve a bi-level optimization as below, KKT optimality condition of the lower level must be added to the higher level problem [32]–[34].

$$\begin{aligned}
&\max \quad F(x_i) \quad , \quad i = 1, \ldots, N \\
&S.T. \\
&\max \quad f(x_i) \\
&\sum_{i=1}^{N} a_i x_i = b \rightarrow \mu_i^{eq} \\
&x_i^{min} < x_i \rightarrow \mu_i^{-} \\
&x_i < x_i^{max} \rightarrow \mu_i^{+} \\
\\
&\downarrow KKt \\
\\
&\nabla_{x_i} L = \frac{\partial f(x_i)}{\partial x_i} - \mu_i^{eq} a_i + \mu_i^{+} - \mu_i^{-} = 0 \\
&\sum_{i=1}^{N} a_i x_i = b \\
&\mu_i^{-}(x_i - x_i^{min}) = 0 \\
&x_i + s_i = x_i^{max} \\
&\mu_i^{+} s_i = 0 \\
&\mu_i^{-}, \mu_i^{+}, \mu_i^{eq}, s_i \geq 0
\end{aligned} \qquad (66)$$